\documentclass[11pt]{article}
\usepackage{fullpage}
\usepackage[latin9]{inputenc}
\usepackage{amsmath}
\usepackage{epsfig}
\usepackage{graphicx}
\usepackage{graphics}
\usepackage[square, comma, numbers,sort&compress, merge]{natbib}
\usepackage[english]{babel}
\usepackage{sectsty}
\sectionfont{\large}
\subsectionfont{\normalsize}
\usepackage{subfigure}
\usepackage{hyperref}
\usepackage{acronym}
\usepackage{amssymb}
\begin{document}
\newcommand{\dirac}{\overline}
\title{\Large \bf \boldmath Enhancement of $h \rightarrow \gamma \gamma$ in the Two Higgs Doublet Model Type I \unboldmath \vspace{0.7cm} }
\author{\normalsize Paul Posch\\  \newline
 \small{Faculty of Physics, University of Vienna, 
 Boltzmanngasse 5, A-1090 Wien, Austria }}
\date{}
\maketitle
\newcommand{\eh}[1]{\mathrm{#1}}
\newcommand{\e}{\epsilon}
\newcommand{\Div}{\text{Div}}
\begin{abstract}
We show that in the \ac{THDM} Type I the partial decay width $\Gamma(h\rightarrow \gamma \gamma)$ can be considerably larger than in Type II, due to light charged scalars with $M_{H^\pm}\approx 100 \text{ GeV}$, which are not yet excluded in Type I. 
 A possible enhancement of the width compared to the SM is analyzed for different Higgs potentials, subject to constraints from tree-level unitarity, vacuum stability and electroweak precision tests.\vspace{0.7cm}
\end{abstract}
  \acrodef{ATLAS}{A Toroidal LHC ApparatuS}
  \acrodef{BSM}{Beyond the Standard Model} 
  \acrodef{CMS}{Compact Muon Solenoid}
  \acrodef{CP}{Charge Parity}
  \acrodef{DM}{Dark Matter}
  \acrodef{FCNC}{Flavour Changing Neutral Currents} 
  \acrodef{IDM}{Inert Doublet Model}
  \acrodef{ILC}{International Linear Collider} 
  \acrodef{KamLAND}{Kamioka Liquid Scintillator Antineutrino Detector}
  \acrodef{LEP}{Large Electron Positron Collider} 
  \acrodef{LHC}{Large Hadron Collider}
  \acrodef{LO}{Leading Order} 
  \acrodef{MSSM}{Minimal Supersymmetric Standard Model}
  \acrodef{NLO}{Next to Leading Order}
  \acrodef{NNLO}{Next to Next to Leading Order}
  \acrodef{PDF}{Parton Distribution Function}
  \acrodef{QCD}{Quantum chromodynamics}
  \acrodef{QED}{Quantum electrodynamics}
  \acrodef{SLC}{Stanford Linear Collider}
  \acrodef{SM}{Standard Model}
  \acrodef{SNO}{Sudbury Neutrino Observatory}
  \acrodef{THDM}{Two Higgs Doublet Model}
  \acrodef{vev}{vacuum expectation value}
\section{Introduction}
While most aspects of the Standard Model (SM) have been confirmed by experiment, the electroweak symmetry breaking mechanism has yet to be established. The SM realizes symmetry breaking with a single Higgs doublet, giving rise to only one physical neutral scalar. 

In the \ac{THDM} one increases the scalar particle content by adding a second doublet, resulting in 5 Higgs bosons ($h$, $H$, $A$, $H^\pm$). The most general version of this model is however problematic, because it induces \ac{FCNC} in the Yukawa sector. One usually introduces discrete symmetries to solve this problem, leading to the Type I and Type II versions of the model \citep{cpstudies}. Having the Yukawa structure of the \ac{MSSM}, the Type II model has been studied extensively in the literature. Despite its interesting phenomenology, the Type I model has received less attention in the literature and has mostly been studied in the fermiophobic limit \cite{Akeroyd:1995hg,  *Brucher:1999tx, Akeroyd:1998ui, Barroso:1999bf}. 

In the Type II model one can derive a strong exclusion limit for the mass of the charged scalar $M_{H^\pm} \gtrsim 300 \text{ GeV}$ independently of $\tan\beta$ (ratio of scalar vacuum expectation values) because of indirect constraints from $B\rightarrow X_s \gamma$ \citep{Ciuchini:1997, *Borzumati:1998add, *Gambino:2001ew}. Due to the different Yukawa interactions, the situation for $M_{H^\pm}$ is different in the Type I model. The strongest lower bound arises from searches at LEP, still allowing for a light charged Higgs with $M_{H^\pm} \gtrsim 90\text{ GeV}$ \citep{Abdallah:2003wd}. 
As most of the early searches at the \ac{LHC} for Higgs bosons will focus on a SM-like light neutral Higgs, it is important to study how much deviation from the SM one can expect for the \ac{THDM}. One suitable decay channel to search for such deviations is $\hat{H} \rightarrow \gamma \gamma$, 
for which $\sigma(p p \rightarrow \hat{H} ) {\cal B}(\hat{H}\rightarrow \gamma \gamma)$ may be measured with a relative error of 10 to 15\% for $ 100 \text{ fb}^{-1}$ integrated luminosity for a SM Higgs $ \hat{H}$ with $110 \text{ GeV} \lesssim M_{\hat{H}} \lesssim 150 \text{ GeV}$ \citep{Zeppenfeld:2000td}.

The decay $h \rightarrow \gamma \gamma$ was studied previously in Type II in Refs. \citep{Arhrib:2003,Ginzburg:2001wj}, which showed that an enhancement of the partial decay width $\Gamma(h \rightarrow \gamma \gamma)$ by around 25\% may occur.

For Type I, this decay was studied in the fermiophobic limit in Refs. \citep{Barroso:1999bf,Akeroyd:2007yh}. In this limit, ${\cal B}(h \rightarrow \gamma \gamma)$ might get considerably enhanced because $h \rightarrow b \dirac{b}$ is loop-suppressed, resulting in a smaller total width of the Higgs boson $h$ compared to the width of the SM Higgs.
Such fermiophobic Higgs bosons might hence lead to improved detection scenarios via $h \rightarrow \gamma \gamma$ at the LHC \citep{Akeroyd:1998ui}. 

On the other hand, cancellation effects may lead to $\Gamma(h \rightarrow \gamma \gamma) \approx 0$, which would render Higgs detection via this decay impossible \citep{Phalen:2006ga}.

In our study, we are interested in enhancement effects of $\Gamma(h \rightarrow \gamma \gamma)$ itself. While $\Gamma(h \rightarrow \gamma \gamma)$ can be measured very precisely at the \ac{ILC}, enhancement effects of it may also be visible early on at the LHC if the Higgs production in the THDM is not suppressed compared to the SM.
Recently, such effects were studied in the context of Higgs production at a photon collider \citep{Bernal:2009rk}.
In our study, we analyze the effect of light charged Higgs bosons with $M_{H^\pm}\approx 100 \text{ GeV}$ for $\Gamma(h \rightarrow \gamma \gamma)$ in the \ac{THDM} Type I in the range $110 \text{ GeV} \lesssim M_h \lesssim 150 \text{ GeV}$, constraining the parameter space with constraints from vacuum stability, unitarity, and electroweak precision tests. The constraints considered are more restrictive than the ones of Ref. \citep{Bernal:2009rk}, and therefore the possible enhancement we find is considerably smaller but yet still enough to be seen at the LHC.

\section{Background}
After introducing the $Z_2$ symmetry $\Phi_1 \rightarrow \Phi_1$, $\Phi_2 \rightarrow - \Phi_2$ for the two Higgs doublets $\Phi_1$, $\Phi_2$ and allowing for a soft breaking term, we get the potential
\begin{align}
          V &=    m_{11}^2 \Phi_1^{\dagger}\Phi_1+ m_{22}^2 \Phi_2^{\dagger}\Phi_2 - \left[m_{12}^2\Phi_1^{\dagger} \Phi_2 + h.c.\right] \nonumber \\
                    & +\dfrac{1}{2} \lambda_1 \left( \Phi_1^{\dagger}\Phi_1 \right)^2 + \dfrac{1}{2}\lambda_1 \left( \Phi_2^{\dagger}\Phi_2 \right)^2
                      +\lambda_3 \left(\Phi_1^{\dagger}\Phi_1 \right)\left(\Phi_2^{\dagger}\Phi_2\right)+ \lambda_4 \left(\Phi_1^{\dagger}\Phi_2 \right)\left( \Phi_2^{\dagger}\Phi_1 \right) \nonumber \\
		      & \left\{+ \dfrac{1}{2}\lambda_5 \left(\Phi_1^{\dagger}\Phi_2\right)^2 + h.c.\right\}\label{eq:potential} \ .
\end{align}
As we will not consider CP violation, all parameters are assumed to be real.
We define the special cases 
\begin{itemize}
\item $V_A$ with $m^2_{12}=0$
\item $V_B$ with $\lambda_5=0$. 
\end{itemize}
The naming convention follows Ref. \citep{Barroso:2007}.
After minimization of the potential \label{eq:z2wv-potential}, one introduces the vevs of the Higgs doublets $\Phi_1$, $\Phi_2$:
\begin{align}
          \langle \Phi_1 \rangle = \dfrac{1}{\sqrt{2}} \left(
	  \begin{array}{c}
	     0 \\
	      v_1
	    \end{array}\right)
	    \ , \qquad \qquad
	  \langle  \Phi_2 \rangle = \dfrac{1}{\sqrt{2}} 
	    \left(
	    \begin{array}{c}
	       0 \\
	        v_2
	      \end{array}\right)\ , \qquad \tan\beta \equiv \dfrac{v_2}{v_1}\ . \label{eq:vevs}
\end{align}
As we only consider minima which do not violate CP, we assume that both $v_1$ and $v_2$ are positive real parameters. Only the potential $V_A$ allows for either $v_1$ or $v_2$ to be exactly zero. In this case, the Higgs doublet with the non-zero vacuum expectation value must couple to the SM particles. This version is called the \ac{IDM} in the literature, and we will take $v_1=0$ in this model \citep{IDM}. As the $Z_2$ symmetry is unbroken in this model, the lightest particle of $\Phi_1$ cannot decay and will contribute to Dark Matter.

In all versions of the model, $v= \sqrt{ v_1^2+v_2^2 } \approx 246 \text{ GeV}$ to yield the correct masses of $M_Z$ and $M_W$. As all parameters are real we can diagonalize the mass matrix of the two CP-even Higgs bosons with a single parameter $\cos\alpha$ (see \citep{cpstudies} for more details). 

For the potential $V$ we can express all parameters but $m_{12}^2$ with the four scalar masses ($M_h$, $M_H$, $M_A$, $M_{H^\pm}$), $\tan\beta$ and $\cos\alpha$. As the potentials $V_A$ and $V_B$ have one parameter less, they can be completely expressed with the Higgs masses, $\tan\beta$ and $\cos\alpha$. 
In the \ac{IDM} there are no mixing angles, and 2 parameters ($\lambda_1$ and $m_{11}^2$) will remain unexpressed in this model.

The Yukawa interaction in the Type I model are defined so that only $\Phi_2$ couples to the fermions. In our convention, the coupling $h \dirac{b} b \propto \cos\alpha/\sin\beta$, where $h$ denotes the lightest CP-even Higgs boson (or the SM-like Higgs boson in the case of the IDM).

\section{The constraints}

\subsection{Vacuum stability}
If we want our vacuum to be not only a local, but also the global minimum, one has to impose the following conditions on the parameters of the potential in \eqref{eq:potential} \citep{Nie:1998yn, *Kanemura:1999xf, *Ferreira}:
\begin{align}
  \lambda_1 \text{, } \lambda_2 \geq 0 \text{ and } -(\lambda_1 \lambda_2)^{1/2} <  \lambda_3 \nonumber\\
  -(\lambda_1\lambda_2)^{1/2} < \lambda_3 + \lambda_4 - \vert \lambda_5 \vert \label{uneq:vacuumstability-ususal}\ .
\end{align}
These conditions yield important constraints on the Higgs masses and mixing angles.
\subsection{Unitarity}
Unitarity constraints in the \ac{THDM} for the potential $V$ were calculated in Ref. \citep{Kanemura:1993hm, *Akeroyd:2000}. 
In this analysis various processes were used to constrain quartic couplings at the tree level. These lead to a set of unitarity constrained parameters $a_\pm, b_\pm , c_\pm , d_\pm , f_\pm , e_{1,2} , 
f_{1,2} , p_1$, whose absolute values must be smaller than $8\pi$. 
These depend on the parameters of the potential, which in the case of $V$ are the 4 Higgs masses, $\tan\beta$, $\cos\alpha$ and $m_{12}^2$\footnote{For a translation of the parameters used in Ref. \citep{Kanemura:1993hm, *Akeroyd:2000} to the ones employed in Eq. \eqref{eq:potential} see Ref. \citep{Gunion:2002zf}}. For the potential $V_A$ one simply sets $m_{12}^2=0$, and for $V_B$ $m_{12}^2=M_A^2 \sin \beta \cos\beta$. In the IDM the above constraints depend on the 4 Higgs masses, $\lambda_1$ and $m_{11}^2$.

\subsection{\texorpdfstring{\boldmath $\Delta r$ \unboldmath}{Delta r}}
Another powerful constraint comes from the well measured constant $G_F=1.16637(1)\times 10^{-5}\text{GeV}^{-2}$, which is defined via the muon decay $\mu \rightarrow e\ \nu_\mu \dirac{\nu}_e (\gamma)$ in the effective Fermi theory. Calculating these decays in the \ac{THDM} one can relate $G_F$ to the self energies of the vector bosons \citep{Bohm:2001yx}:
\begin{align}
  \dfrac{G_F}{\sqrt{2}} = \dfrac{\alpha\ \pi}{2 s_W^2 m_W^2}(1+\Delta r)\ . \label{eq:sirlin}
\end{align}
Here $s_W$ is the sine of the weak mixing angle $s_W= \sin \theta_W$ with
\begin{align}
  \sin^2 \theta_W =1 - \dfrac{m^2_W}{m^2_Z}\ .
\end{align}
The quantity $\Delta r$ is then separated into the finite quantities
\begin{align}
  \Delta r = \Delta \alpha - \dfrac{c^2_W}{s^2_W}\Delta \rho + \Delta
  r_{\text{rem}} \label{eq:Deltar}\ , 
\end{align}
\begin{align}
  &\Delta \rho \equiv \dfrac{\Sigma^{ZZ}(0)}{m^2_Z} - \dfrac{\Sigma^{WW}(0)}{m^2_W} - 2 \dfrac{s_W}{c_W}\dfrac{\Sigma^{AZ}(0)}{m^2_Z} , \qquad  \Delta \alpha \equiv -Re\Pi^{AA}(m^2_Z) + \Pi^{AA}(0) \ , \\
  &\Delta r_{\text{rem}} \equiv \left(1-\dfrac{c^2_W}{s^2_W}\right)\dfrac{\Sigma^{WW}(0) - Re\Sigma^{WW}(m^2_W)}{m^2_W}
   + Re\Pi^{AA}(m^2_Z) \nonumber \\
   &\phantom{\Delta r_{\text{rem}} \equiv}
   +\dfrac{c^2_W}{s^2_W}\dfrac{\Sigma^{ZZ}(0) -
   Re\Sigma^{ZZ}(m^2_Z)}{m^2_Z}+\dfrac{\alpha}{4\pi s^2_W}\left(6 +
   \dfrac{7-4 s^2_W}{2s^2_W}\log c^2_W \right) \ .
 \end{align}
 Here $\Sigma^{WW}$, $\Sigma^{Z Z}$, $\Sigma^{A Z}$ denote the transversal parts of the self-energies of the $W$ and $Z$ boson. $\Pi^{AA}$ is defined as
\begin{align}
  \Pi^{AA}(k^2)\equiv \dfrac{\partial \Sigma^{AA}(k^2)}{\partial\ k^2} \ ,
 \end{align}
 where $\Sigma^{AA}$ is the transversal self-energy of the photon. $\Delta \rho$ and $\Delta r_{rem}$ at the 1-loop level were calculated in FeynArts/FormCalc \citep{feynarts,formcalc} and then evaluated with LoopTools \citep{formcalc, vanOldenborgh:1989wn}.
 We take $ \Delta \alpha(m_Z^2)= 0.0594(5)$, where most of the uncertainties come from the hadronic contributions \citep{Hollik:2003}.
 Note that due to the Appelquist-Carazzone theorem all heavy particles like the top and the Higgs bosons decouple from $\Delta \alpha$ \citep{Appelquist:1974}.
 On the contrary, $\Delta \rho$ and $\Delta r_{rem}$ are sensitive to the masses of the Higgs bosons, the W and Z bosons and the top quark.
 
 As $\Delta \rho$ depends quadratically on the top mass, while $\Delta r_{rem}$ only depends logarithmically on it, we include 3-loop QCD corrections only for the $\Delta \rho$ parameter, which were calculated in Ref. \citep{Chetyrkin:1995}.
 If one takes into account 2-loop contributions in the SM, the value of $\Delta r$ shifts up by $\approx 0.005$ \citep{MW-SM-prediction:2003}. We take this as a rough estimate for the error arising from neglecting the 2-loop contributions in the \ac{THDM}. 

On the other hand, one can simply insert the precisely measured values $m_Z= 91.1876(21) \text{ GeV}$, $\alpha(0) = 1/137.03599976(50)$ and $m_W=80.398(25) \text{ GeV}$ into \eqref{eq:sirlin}, which then yields an experimental prediction for $\Delta r$:
\begin{equation}\label{eq:rexp}
  \Delta r_{exp} = 0.0343 \pm 0.0020 \ .
\end{equation}
Combining the uncertainty from experiment at 2 $\sigma$ with the theoretical error due to neglecting the 2-loop contributions, we can exclude $\Delta r < 0.0253$ and $\Delta r > 0.0433$. At the one-loop level $\Delta r$ is only sensitive to 5 unknown parameters, $\sin(\beta - \alpha)$, $M_{H^\pm}$, $M_A$, $M_h$, $M_H$. For the top quark we use the value $m_t=171.2(2.3)$ \citep{pdg:2008}.

 \subsection{Z decays}
Yet another observable which has been determined with great precision at LEP and SLAC is the hadronic branching ratio of Z to $b \dirac{b}$ pairs:
 \begin{align}
   R_b \equiv \dfrac{\Gamma(Z\rightarrow b\bar{b})}{\Gamma(Z\rightarrow \text{hadrons})}.
 \end{align}
 A recent data analysis in Ref. \citep{Abe:2004} showed that
 \begin{align}
   R_b=0.2163 \pm 0.0007 \ ,
   \label{eq:Rbexp}
 \end{align}
 which is in good agreement with the SM, which yields $R_b^{SM} = 0.2158$  \citep{Field:1997}.
 The ratio was calculated in the \ac{THDM} in Refs. \citep{Denner:1991} and \citep{Haber:1999}. For Type I, one finds that the contribution of the charged Higgs bosons dominates the contribution of the neutral Higgs bosons, and therefore we can neglect the latter. One can hence derive a lower bound for $\tan\beta$ that depends on the charged Higgs mass. Using the appropriate formula of \citep{Haber:1999} one finds that for $m_t=171.2 \text{ GeV}$   
\begin{align}
&\tan\beta \lesssim 2 \quad \text{excluded for}\quad M_{H^\pm} \approx 100
\text{ GeV} \label{eq:lowtanbetaexclusion-Rb} 
\end{align}
at 2 $\sigma$.
Increasing the Higgs mass the bound drops (e.g. for $M_{H^\pm} \approx 500 \text{ GeV}$ one can exclude $\tan\beta \lesssim 1$). One can find similar bounds by considering $B\rightarrow X_s \gamma$ \citep{Aoki:2009ha,*Su:2009fz, *Logan:2009uf} \footnote{The exact value of the bound is less important for our study. What is important, is that we get a lower bound for $\tan\beta$ which is well above $1$ for $M_{H^\pm} \approx 100\text{ GeV}$.}.

\section{\texorpdfstring{Analysis of \boldmath  $h \rightarrow \gamma\gamma$ \unboldmath}{Analysis of h decaying into two photons}}
Having introduced the constraints, we will apply them to constrain possible enhancement effects in $\Gamma(h \rightarrow \gamma \gamma)$.
The partial decay width $\Gamma(h \rightarrow \gamma \gamma)$ is loop induced, and can be easily calculated in FeynArts. The 1-loop analytical result is well known (see \citep{mssm-higgs-review} and references therein):
\begin{align}
  \Gamma(h\rightarrow \gamma \gamma)  = \dfrac{G_{F}\alpha^2 M^3_h}{128 \sqrt{2} \pi^3} \bigg\vert &\sum_{f \in \{t,b\}}g_fQ_f^2N_c A_{1/2}(\tau_f) + g_W A_1(\tau_W) + g_h A_0(\tau_{H^\pm}) \bigg \vert^2 , \label{eq:Gammahphopho}
\end{align}
where  
\begin{align}
g_W=\sin(\beta-\alpha)\ , \qquad g_f=\dfrac{\cos\alpha}{\sin\beta} , \qquad 
g_h=-\dfrac{m_W}{g m_{H^\pm}^2}g_{h H^+ H^-}\ , 
\end{align}
with the contribution of the other light fermions neglected. $g_W$ and $g_f$ are the trilinear couplings of $h$ to the $W$ gauge bosons and to the fermions normalized to the ones of the SM.  $g_{h H^+ H^-}$ is the coupling which appears in the Lagrangian, ${\cal L}=g_{h H^+ H^-} h H^+ H^- \cdots$.
$\tau_i= \tfrac{M_h^2}{4 m_i^2}$, and the form factors $A_i$ are defined as
\begin{align}
A_0(\tau)&=-[\tau-f(\tau)]\tau^{-2}\nonumber \ , \\
A_{1/2}(\tau)&=2[\tau + (\tau - 1)f(\tau)]\tau^{-2} \nonumber \ , \\
A_{1}(\tau)&=-[2\tau^2+3\tau+3(2\tau - 1)f(\tau)]\tau^{-2} \ ,
\end{align}
and 
\begin{eqnarray}
  f(\tau) =\left\{  
  \begin{array}{ll}  \displaystyle
    \arcsin^2\sqrt{\tau} & \tau\leq 1 \\
    \displaystyle -\frac{1}{4}\left[ \log\frac{1+\sqrt{1-\tau^{-1}}}
    {1-\sqrt{1-\tau^{-1}}}-i\pi \right]^2 \hspace{0.5cm} & \tau>1
  \end{array} \right. .
\end{eqnarray}
Note that $g_f$ is the same for $b$ and $t$ quarks in Type I. To get the result in the SM, one simply sets $g_f=g_W=1$ and $g_h=0$. The result for Type II is almost the same, only $g_t=\cos\alpha/\sin\beta$ and $g_b=\sin\alpha/\cos\beta$. If the coupling $h \dirac{b} b$ is not enhanced, the contribution of the $b$ quarks is negligible and the result is approximately the same for both types. The main difference between the Type I and Type II models hence comes from the different constraints available, particularly the light charged Higgs bosons, which are not yet excluded in Type I. As we only consider Type I, the contribution of the $b$ quarks (and the other fermions) will be neglected in the following discussion.

In the parameter space region where $110 \text{ GeV} \leq M_h \leq 150 \text{ GeV}$ and $M_{H^\pm}= 100 \text{ GeV}$, we can see that the contribution of the W bosons is by far dominant, being around $ -10 \lesssim A_1 \lesssim -8$, while $0.4 \lesssim A_0 \lesssim 0.5$ and $A_{1/2} \approx 1.4$\footnote{For plots of these functions, see \citep{mssm-higgs-review}}. Due to the fact that $g_f$ may only rise for small $\tan\beta$, which is restricted by the $R_b$ constraint,
the most significant enhancement arises from $g_h$. 

For a useful comparison we define the ratio $R_{\gamma\gamma}$, the partial decay width normalized to its value in the SM:
\begin{align}
  R_{\gamma\gamma}=\dfrac{\phantom{a}\Gamma(h\rightarrow \gamma \gamma)_{\phantom{SM}}}{\phantom{a}\Gamma(h\rightarrow \gamma \gamma)_{SM}} \label{eq:Rgaga} .
\end{align}
We then maximize this ratio subject to the constraints introduced in the previous sections. For the maximization procedure we assume $110 \text{ GeV} \leq M_h \leq 150 \text{ GeV}$, and $M_{H^\pm}= 100 \text{ GeV}$ for the mass of the charged scalars. 
For the masses of the other Higgs bosons we assume $M_h \lesssim M_H \lesssim 1 \text{ TeV}$ (except in the IDM, where we may also have $M_h > M_H$), and $100 \text{ GeV} \lesssim M_{A} \lesssim 1 \text{ TeV}$.
The maximization was done in Mathematica 6, using its three different maximization algorithms.
While these algorithms do not prove that a certain point is a global maximum, they easily find the regions where the width is enhanced.

\subsection{IDM}
\begin{figure}[htbp!]
\subfigure[\ $R_{\gamma\gamma}$ for $\lambda_{1}=8.2$. The allowed region increases slightly for higher values of $m_h$. ]
{
\includegraphics[scale=0.65]{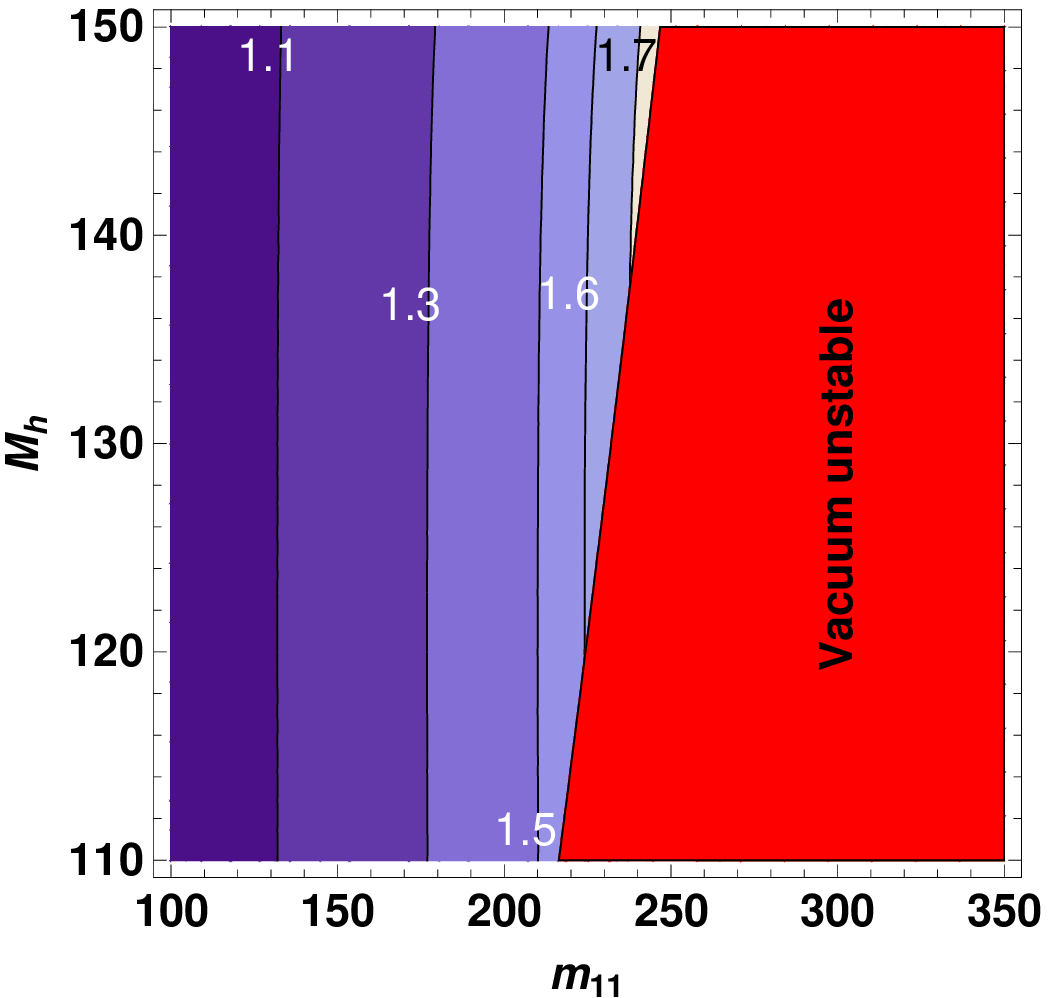}
\label{fig:GammaIDMm11max}
}
\hspace{1cm}
\subfigure[\ $R_{\gamma\gamma}$ for $\lambda_{1}=8.355$. We see that unitarity constraints forbid larger values of $\lambda_{1}$ for high values of $m_{11}$.]
{
\includegraphics[scale=0.65]{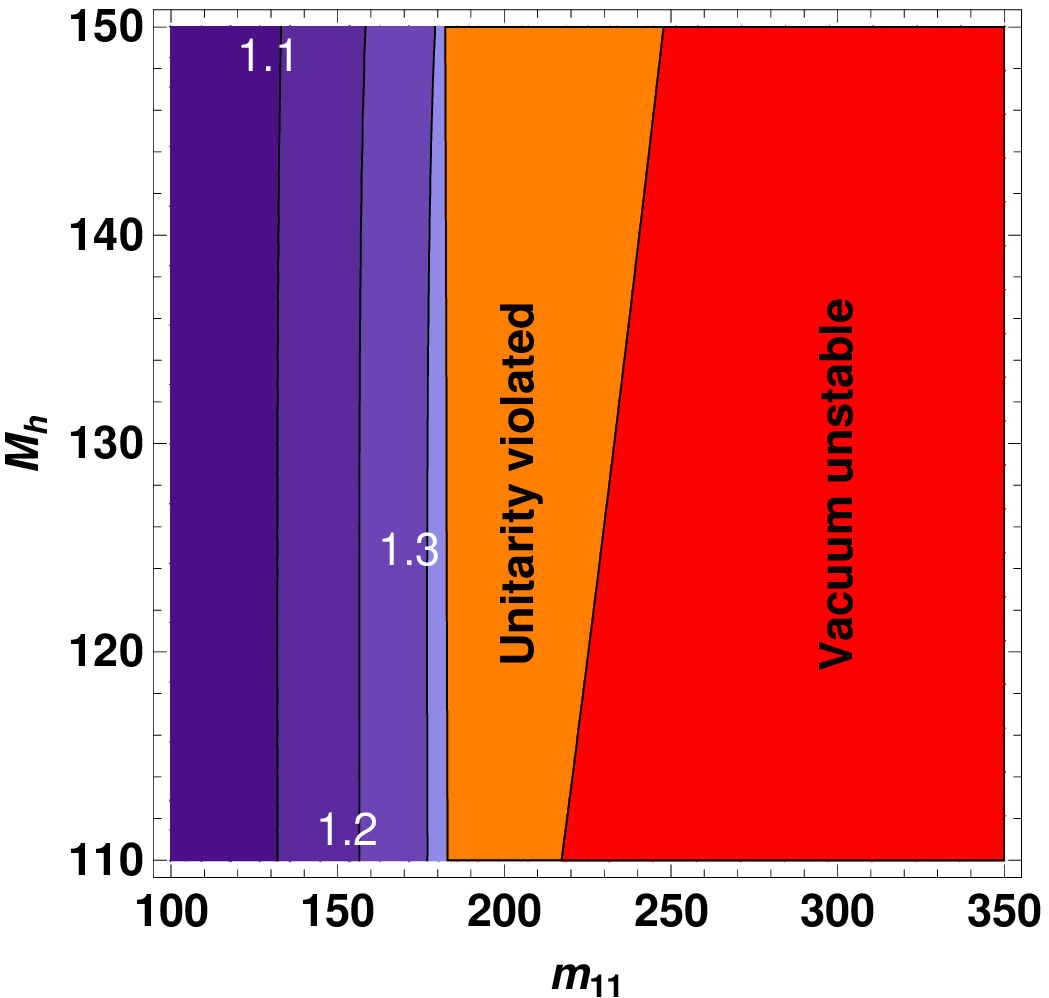}
\label{fig:GammaIDMm11constrained}
}
\caption{Curves of constant value of $R_{\gamma \gamma}$ are plotted in the ($M_h$, $m_{11}$) plane for the IDM.
The exclusion regions were calculated for  $M_{H^\pm}= 100 \text{ GeV}$, $M_H=90 \text{ GeV}$ and $M_A=120 \text{ GeV}$. $\Delta r\approx 0.038$ in both regions displayed.
 }\label{fig:cpbs}
\end{figure}

As the Higgs boson $h$ of the IDM behaves exactly like the SM Higgs when coupling to SM particles, we have $g_W=g_f=1$ (which corresponds to $\alpha=0$, $\beta=\pi/2$). For the trilinear coupling to the charged Higgs bosons we get $g_{h H^+ H^-} = - \lambda_3 v$, which results in 
\begin{align}
  g_h=\dfrac{M^2_{H^\pm} - m_{11}^2}{M_{H^\pm}^2} \ .
\end{align}
As $A_1$ is negative and $A_0$ positive, we must have $m_{11}^2 \gg M_{H^\pm}^2$ to get constructive interference. This means we must have $\lambda_3 < 0 $ and $-\lambda_3 \gg 1$, which is constrained by the vacuum stability conditions in Eq. \eqref{uneq:vacuumstability-ususal}. If $m_{11}^2$ i.e. $\lambda_3$ is large, $\lambda_1$ must also be large to be compatible with the constraints from vacuum stability ($\lambda_2$ cannot compensate for $\lambda_1$, because we have $\lambda_2 = M_h^2/v^2$ and $M_h$ is fixed in our discussion). Furthermore, large values of $\lambda_1$ and $\lambda_3$ are constrained by unitarity, especially $\vert a_+ \vert \leq 8\pi$. We hence find maximal enhancement to be around $1.6 \lesssim R_{\gamma \gamma} \lesssim 1.8$ in the region where $110 \text{ GeV} \leq M_h \leq 150 \text{ GeV}$ for light charged scalars with $M_{H^\pm}=100 \text{ GeV}$. As $M_{H^\pm}^2 \lesssim v^2$, increasing $M_{H^\pm}$ will not decrease $a_+$ significantly to allow for larger $m_{11}^2$, and only results in a smaller $g_h$ and therefore an overall smaller enhancement.
We plot a possible enhancement region in Fig. \ref{fig:GammaIDMm11max}. As shown in Fig. \ref{fig:GammaIDMm11constrained}, further increasing $\lambda_1$ weakens the vacuum stability constraint, but increases the region forbidden by the unitarity constraints.

If $m_{11}^2$ is negative, the contributions interfere destructively. The contribution of the charged Higgs bosons must hence get more than twice the contribution of the W bosons to get enhanced. Using $\vert a_+\vert \leq 8\pi$ one can derive the rough bound $m_{11} \lesssim  600 \text{ GeV}$, which yields $R_{\gamma\gamma} \lesssim 0.9$, and therefore no enhancement is possible in this case.
\newpage
\subsection{\texorpdfstring{\boldmath $V_A$ \unboldmath}{VA}}

For the potential $V_A$ (with both $v_1,v_2 \neq 0$), we have 
\begin{align}
g_{h} = g_1 + g_2 \ ,
\end{align}
where
\begin{align}
&g_1 = s_{\beta-\alpha}\left(1-\dfrac{M^2_{h}}{2 M^2_{H^\pm}} \right) \nonumber \ , \\
&g_2 = \dfrac{c_{\alpha +\beta}}{s_{2\beta}} \dfrac{M^2_h}{M^2_{H^\pm}} . \label{eq:g1andg2}
\end{align}
As $g_1 \propto g_w$, a significant enhancement is only possible for large $M_h$. In the region we consider, $ -0.13 \lesssim g_1 \lesssim 0.4 $ and therefore it cannot significantly contribute to an enhancement of $g_h$.

$g_2$ enhances the contribution of the charged Higgs bosons for large $\tan\beta$.
To make the interference effects with the W bosons more obvious, one rewrites Eq. \eqref{eq:Gammahphopho} as
\begin{align}
    \Gamma(h\rightarrow \gamma \gamma)  = \dfrac{G_{F}\alpha^2 M^3_h}{128 \sqrt{2} \pi^3} \bigg\vert c_\alpha C + s_\alpha D  \bigg\vert ^2 , \label{eq:rewrittenGammahphopho}
\end{align}
where 
\begin{align}
  C=s_\beta A_1 + \dfrac{A_0}{s_\beta} \dfrac{M_h^2}{2 M^2_{H^\pm}} + \dfrac{1}{s_\beta} \tilde{A}_{1/2} \ , \qquad 
  D=-\left(c_\beta A_1 + \dfrac{A_0}{c_\beta} \dfrac{M_h^2}{2 M^2_{H^\pm}} \right) \ , \label{eq:contribCD}
\end{align}
with $ \tilde{A}_{1/2}=Q_t^2N_c A_{1/2}(\tau_t)$ and the term proportional to $g_1$ neglected. As both $C$ and $D$ are real for the parameter region considered, maximizing over $c_\alpha$, $s_\alpha$ simply yields
\begin{align}
  \max_{c_\alpha, s_\alpha} \left\{ \Gamma(h\rightarrow \gamma \gamma) \right\}  = \dfrac{G_{F}\alpha^2 M^3_h}{128 \sqrt{2} \pi^3} \bigg\vert C^2 + D^2 \bigg\vert . \label{eq:MaxGammahphopho}
\end{align}
Since $A_0$ and $A_1$ have opposite sign, the contributions of the charged Higgs and the W bosons interfere destructively in both $C$ and $D$. 

Expanding $C$ and $D$ for large $\tan\beta$, one sees that $C$ is of order $1$, while in $D$ the contribution of the W bosons is of order $1/\tan\beta$, and the one of the charged Higgs bosons is of order $\tan\beta$. For moderate values of $\tan\beta$ (i.e. $ 3 \lesssim \tan\beta \lesssim 6$), $C$ is by far dominant, and $D$ is only around 5\% of $C$. 

\begin{figure}[htbp!]
  \subfigure[\ $R_{\gamma \gamma}$ is shown for $M_h=110 \text{ GeV}$. $\Delta r \approx 0.037$ in the region displayed.]
  {
  \includegraphics[scale=0.65]{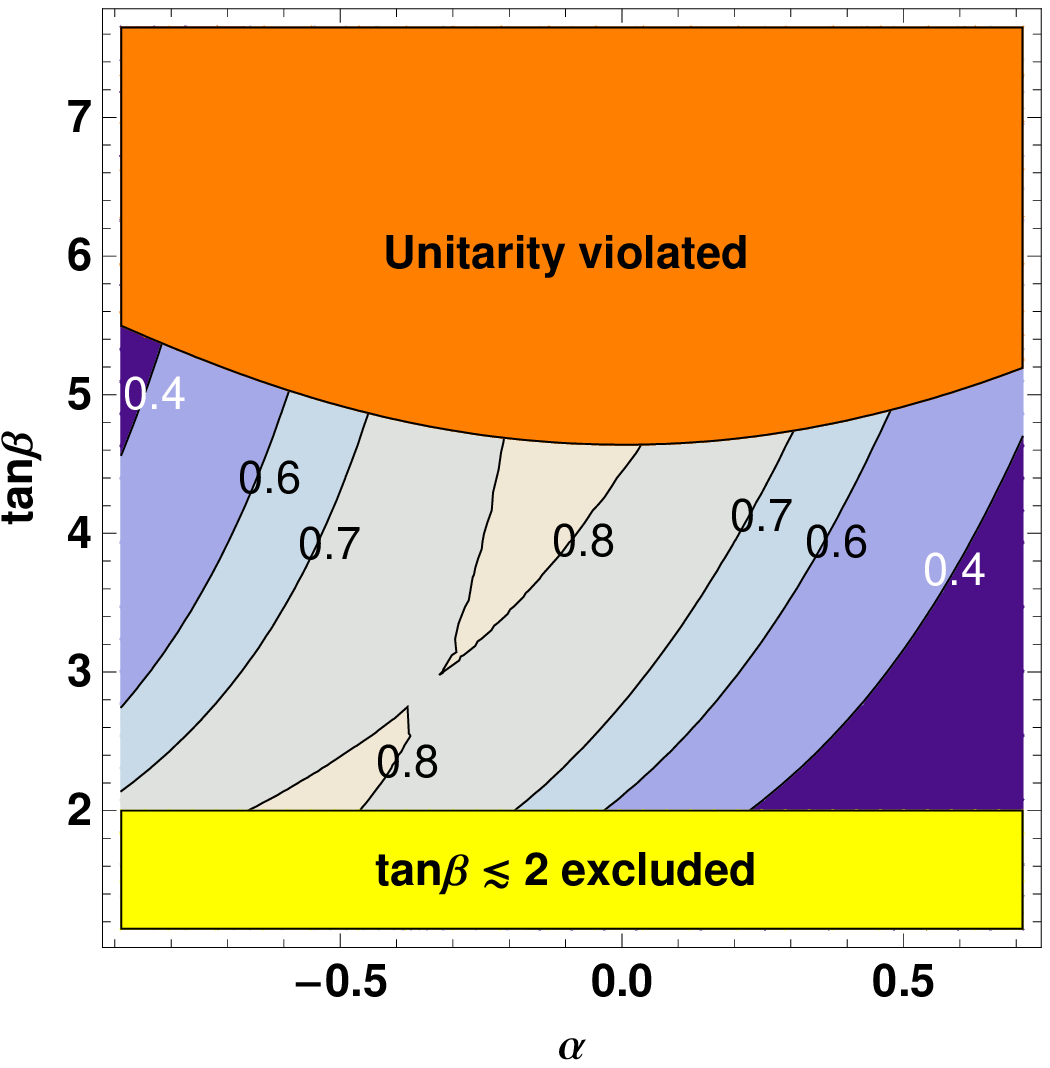}
  \label{fig:GammaVAMHp100Mh110MA100MH150fertig}
  }
  \hspace{1cm}
  \subfigure[\ $R_{\gamma \gamma}$ is shown for $M_h=150 \text{ GeV}$. Increasing $M_h$ decreases $C$ and hence reduces $R_{\gamma \gamma}$. $\Delta r \approx 0.038$ in the displayed region.]
  {
  \includegraphics[scale=0.65]{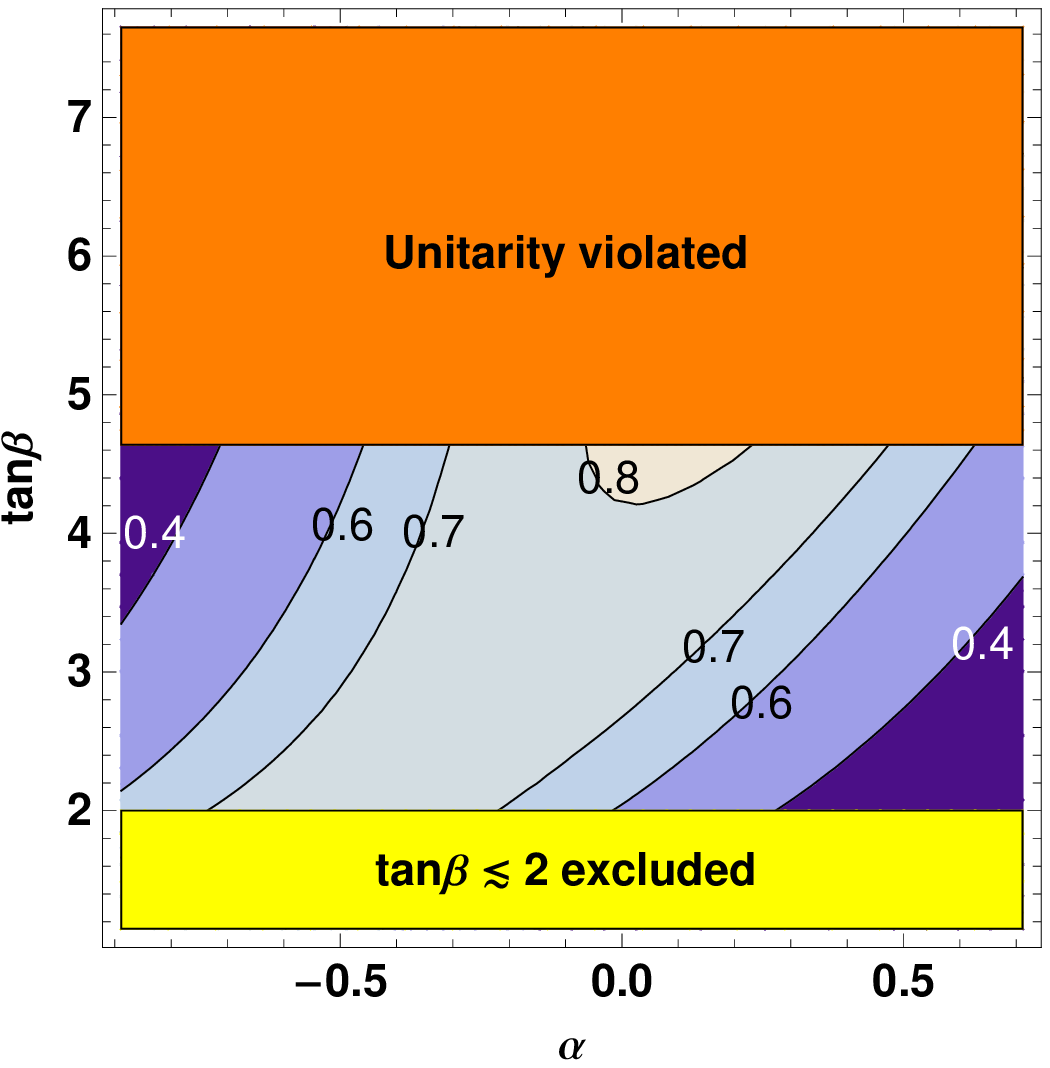}
  \label{fig:GammaVAMHp100Mh150MA100MH150fertig}
  }
  \caption{$R_{\gamma \gamma}$ analyzed for the $V_A$ potential, for $M_{H^\pm}=100 \text{ GeV}$, $M_A=100 \text{ GeV}$, $M_H=150\text{ GeV}$. We see that there is no enhancement, but a reduction compared to the SM. The displayed region is not forbidden by $\Delta r$ or the vacuum stability conditions.}
  \label{fig:VAsuppression}
\end{figure}

When raising $\tan\beta$, $\vert D \vert$ first drops to zero as the two contributions cancel, and then starts to grow again as the contribution of the charged Higgs bosons starts to dominate in $D$. One would hence expect an enhancement from the point on where $D$ compensates for the suppression in $C$, which happens at $\tan\beta \gtrsim 8$.
However, when employing the maximization procedure we find that such large values of $\tan\beta$ are strongly restricted by unitarity (especially $a_+$), and only moderate values of $\tan\beta$ are allowed. The partial decay width hence gets reduced compared to the SM, with typical maximal values $R_{\gamma \gamma} \approx 0.8$ for  $110 \text{ GeV} \lesssim M_h \lesssim 150 \text{ GeV}$ as shown in Fig. \ref{fig:VAsuppression}. As we stay in the region where $C$ dominates, raising $M_h$ raises the negative contribution of the charged Higgs bosons in $C$ and leads to a slightly stronger suppression compared to the SM. 

\subsection{\texorpdfstring{\boldmath $V$ and $V_B$ \unboldmath}{V and VB}}
\begin{figure}[hb!]
\subfigure[\ $R_{\gamma\gamma}$ with $M_H=224\text{ GeV}$, $M_A=102 \text{ GeV}$ and $m_{12}= 50 \text{ GeV}$. $\Delta r\approx 0.035$ in this region and therefore allowed. ]
{
\includegraphics[scale=0.69]{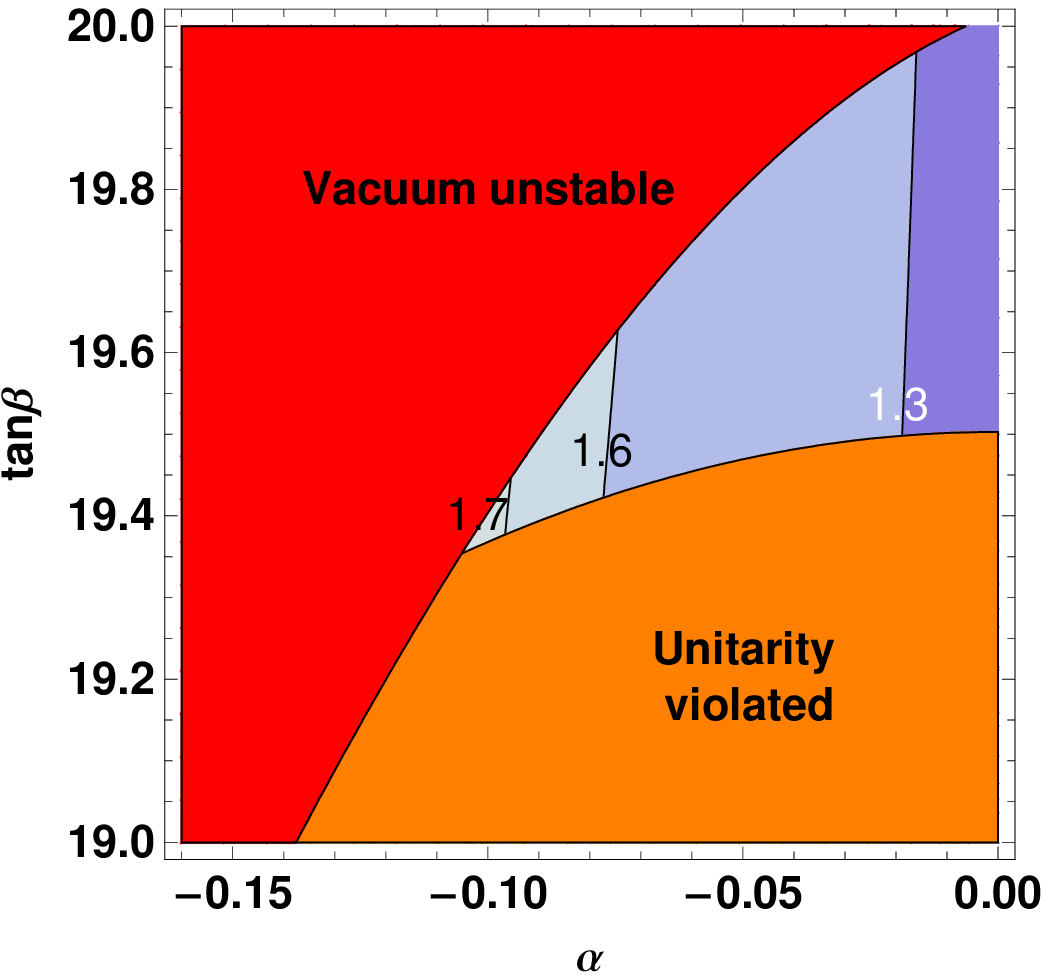}
\label{fig:GammaExVm12MHp100M1250MA102MH224exotic}
}
\hspace{1cm}
\subfigure[\ $R_{\gamma\gamma}$ with $M_H=333\text{ GeV}$, $M_A=111 \text{ GeV}$ and $m_{12}=100 \text{ GeV}$. $\Delta r \approx 0.038$, which is not forbidden at 2 $\sigma$. ]
{
\includegraphics[scale=0.65]{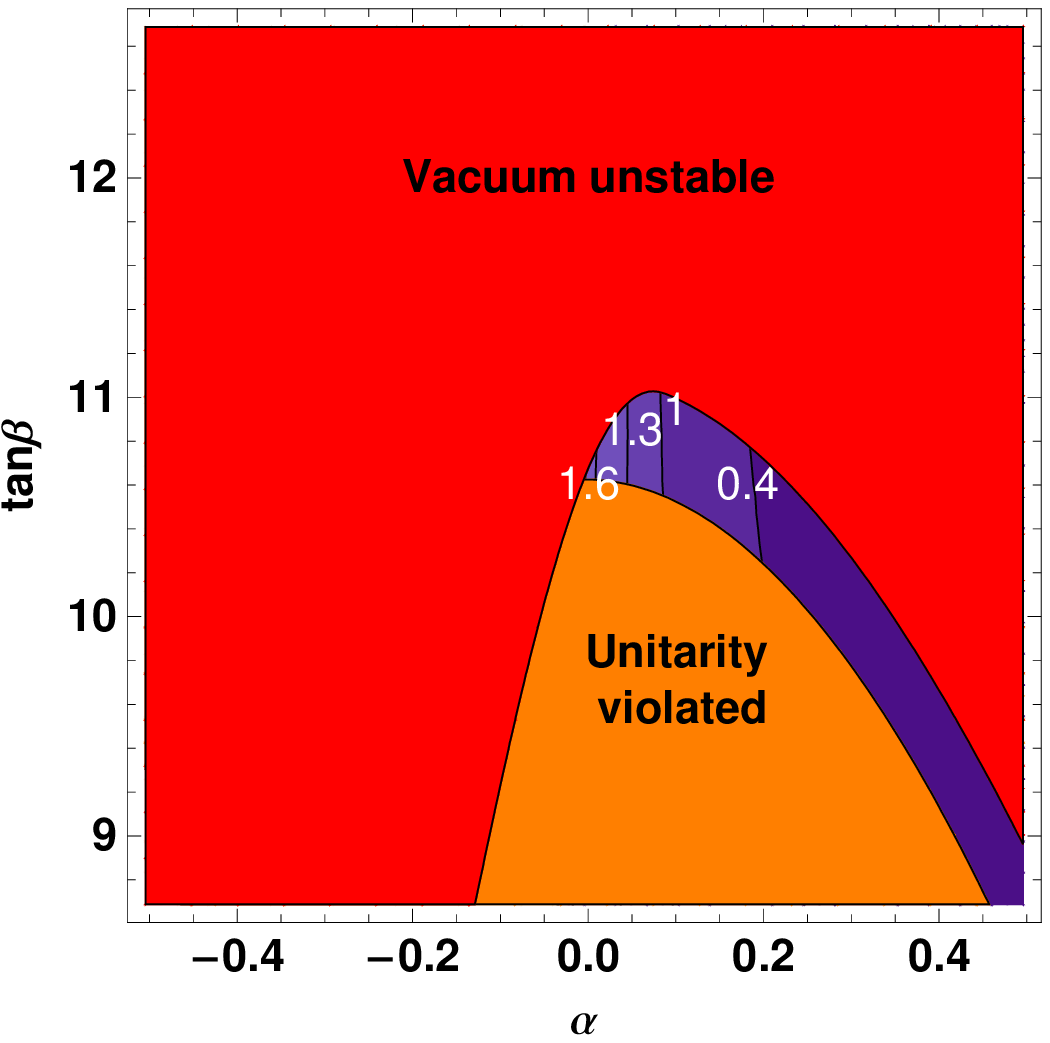}
\label{fig:GammaExVm12MHp100M12100MA111MH333constrained}
}

\caption{$R_{\gamma \gamma}$ analyzed for the $V$ potential \eqref{eq:potential}, for $M_h=130 \text{ GeV}$, $M_{H^\pm}=100 \text{ GeV}$. }
\label{fig:GammaExVm12}
\end{figure}

At last we will study the potential $V$ (and $V_B$), where $m^2_{12}$ is a free parameter. We now have
\begin{align}
g_h=g_1 + g_2 + g_3, 
\end{align}
where $g_1$ and $g_2$ are again the parameters defined in \eqref{eq:g1andg2}. $g_3$ is not fixed via the masses, but is proportional to $m^2_{12}$:
\begin{align}
g_3 = -\dfrac{c_{\alpha +\beta}}{2 s^2_{\beta}c^2_{\beta}} \dfrac{m_{12}^2}{M^2_{H^\pm}} .
\end{align}
$g_3$ has the opposite sign of $g_2$. The interference behavior for $V$ is again illustrated by Eq. \eqref{eq:MaxGammahphopho}, where now in $C$ and $D$
we must replace 
\begin{align}
  \dfrac{M_h^2}{2 M^2_{H^\pm}} \rightarrow \dfrac{M_h^2}{2 M^2_{H^\pm}} - \dfrac{m_{12}^2}{2 s_\beta c_\beta M^2_{H^\pm}} \ .
\end{align}
If $m_{12}^2$ is positive and $-g_3 \gg g_2$, the contributions of the charged Higgs bosons and the W bosons interfere constructively in both $C$ and $D$, and we get an enhancement for large $\tan\beta$ or $m_{12}^2$. 
Applying the maximization procedure, one finds that $R_{\gamma\gamma}\approx 1.7$ can be realized for $110 \text{ GeV} \lesssim M_h \lesssim 150 \text{ GeV}$ while still being compatible with all the constraints considered. Regions where this may be realized are shown in Fig. \ref{fig:GammaExVm12}. 

If $m_{12}^2$ is negative, the situation is similar to the $V_A$ case as we again get destructive interference in both $C$ and $D$. An enhancement is possible if $ m_{12}^2$ is large, or if $\tan\beta$ is large. These regions are strongly constrained by the unitarity constraint $\vert a_+ \vert \leq 8 \pi $, which is more restrictive if $m_{12}^2$ is negative.
As $a_+$ restricts simultaneous growth of $\tan\beta$ and $ m_{12}^2$, one can derive a rough lower bound, which is $-(250 \text{ GeV})^2 \lesssim m_{12}^2$ for $\tan\beta\approx 2$. Even if such low values were allowed, one would only get $R_{\gamma \gamma} \lesssim 0.7 $ and therefore no enhancement compared to the SM. Lower values of $m_{12}^2$ can be attained for smaller values of $\tan\beta$, but we would also have to raise $M_{H^\pm}$ to be compatible with the constraint from $R_b$, and therefore this would not result in an overall enhancement either.

Expanding $a_+$ for large values of $\tan\beta$, one sees that the maximal values of $\tan\beta$ are smaller for non-zero negative values of $m_{12}^2$.
Performing the numerical maximization, we found no enhancement, with the partial width staying below the maximal values attained for $V_A$ in most of the parameter region.

For the potential $V_B$ we have one free parameter less, as $m_{12}^2=M_A^2 s_\beta c_\beta$.
If $M_A^2 \gg M_h^2$ we again have constructive interference in both $C$ and $D$, and the situation is similar to $V$ with large and positive $m_{12}^2$.
The maximal enhancement we found is a bit lower, yielding $R_{\gamma\gamma}\approx 1.6$ for $110 \lesssim M_h \lesssim 150 \text{ GeV}$ \footnote{Plots for $V_B$ can be found in Ref. \citep{Posch}}. Obviously, the parameter space available for such an enhancement is smaller for $V_B$. 
\section{Conclusions}
We discussed a possible enhancement of the partial decay width $\Gamma(h\rightarrow \gamma \gamma)$ in the \ac{THDM} Type I for light charged scalars with $M_{H^\pm}=100 \text{ GeV}$. Unlike in the Type II model, where $M_{H^\pm} \gtrsim 300 \text{ GeV}$ due to constraints from B physics, one cannot exclude light charged scalars in Type I.
We maximized the ratio of the decay width in the \ac{THDM} over the decay width in the SM, subject to constraints from vacuum stability, tree-level unitarity, Z decays into hadrons, and the $\Delta r$ parameter.
The ratio was analyzed for the neutral CP-even Higgs $h$ with  $ 110 \text{ GeV} \lesssim M_h \lesssim 150 \text{ GeV}$ (which is the region where an SM-like light Higgs boson $h$ can be discovered at the \ac{LHC} in $h \rightarrow \gamma \gamma$), and with $M_{H^\pm}=100\text{ GeV}$ for the charged Higgs boson.
The maximal possible enhancement differs for the potentials considered due to different interference scenarios. For the IDM the maximal enhancement was found to be around $+70\%$, for $V_A$ around $-20\%$. The results for $V$ and $V_B$ were rather similar, being around $70\%$ and $60\%$, respectively.
This is larger than what was found for Type II, where for heavy charged Higgs bosons with $M_{H^\pm} \approx 400 \text{ GeV}$ the enhancement was around $ 25\%$ \citep{Arhrib:2003}. Our results for the enhancement of the partial width in Type I differ from Ref. \citep{Bernal:2009rk}, mostly because we used the more restrictive unitarity constraints of Ref. \citep{Kanemura:1993hm, *Akeroyd:2000} and the additional constraint $R_b$ of Ref. \citep{Haber:1999}.

The expected accuracy at the \ac{LHC} for $\sigma(p p \rightarrow \hat{H} ) {\cal B}(\hat{H}\rightarrow \gamma \gamma)$ is around 10 to 15\% for an integrated luminosity of $100 \text{ fb}^{-1}$ \citep{Zeppenfeld:2000td}. If the total decay width and the production cross section in the \ac{THDM} Type I are not too different from the SM, the accuracy should be sufficient to distinguish between the SM and such THDMs.

Combining measurements at the photon collider option of the \ac{ILC} with the $e^+e^-$ collider option, the partial width of a SM Higgs with $M_{\hat{H}}=120 \text{ GeV}$ can be determined with $3\%$ accuracy \citep{Monig:2007py}.
Such measurements, combined with possible direct detection of a charged Higgs may hence be used to distinguish between the various possible potentials of the THDM.
\newpage
\subsubsection*{Acknowledgments}
I am grateful to David Lopez-Val and Brook Thomas for bringing their papers \citep{Bernal:2009rk} and \citep{Phalen:2006ga} to my attention. I am indebted to Shinya Kanemura for pointing out several important references. I want to express my gratitude to Gerhard Ecker for valuable discussions, helpful advices and for carefully reading the manuscript.
\bibliography{biblio}
\bibliographystyle{h-physrev5}
\end{document}